\documentclass[manuscript]{aastex}

\slugcomment{}

\shorttitle{e damping through collisions}
\shortauthors{Matsumoto et al.}

\begin{document}


%\title{Collapsed Cores in Globular Clusters, \\
% Gauge-Boson Couplings, and AAS\TeX\ Examples}

\title{Eccentricity Evolution Through Accretion of Protoplanets}

%\telephone{+8180-5105-9939}

\author{Yuji Matsumoto, \altaffilmark{1}}
\email{yuji.matsumoto@nao.ac.jp}
%\affil{Tokyo Institute of Technology}
%telephone +8180-5105-9939

\author{Makiko Nagasawa\altaffilmark{2}}
%\affil{National Optical Astronomy Observatories, Tucson, AZ 85719}

\and

\author{Shigeru Ida\altaffilmark{3}}
%\affil{Space Telescope Science Institute, Baltimore, MD 21218}

%% are identified by the \altaffilmark after each name. Specify alternate

\altaffiltext{1}{Center for Computational Astrophysics, National Astronomical Observatory of Japan, 2-21-1 Osawa, Mitaka, Tokyo, 181-8588, Japan}
\altaffiltext{2}{International Education and Research Center of Science, Tokyo Institute of Technology, Ookayama, Meguro-ku, Tokyo 152-8551, Japan}
\altaffiltext{3}{Earth-Life Science Institute, Tokyo Institute of Technology, Ookayama, Meguro-ku, Tokyo 152-8550, Japan}


\begin{abstract}

Most of super-Earths detected by the radial velocity (RV) method have significantly smaller eccentricities than the eccentricities corresponding to velocity dispersion equal to their surface escape velocity (``escape eccentricities").
If orbital instability followed by giant impacts among protoplanets that have migrated from outer region is considered, it is usually considered that eccentricities of the merged bodies become comparable to those of orbital crossing bodies, which are excited up to their escape eccentricities by close scattering.
However, the eccentricity evolution in the {\it in situ} accretion model has not been studied in detail.
Here, we investigate the eccentricity evolution through {\it N}-body simulations.
We have found that the merged planets tend to have much smaller eccentricities than the escape eccentricities due to very efficient collision damping.
If the protoplanet orbits are initially well separated and their eccentricities are securely increased, an inner protoplanet collides at its apocenter with an outer protoplanet at its pericenter.
The eccentricity of the merged body is the smallest for such configuration.
Orbital inclinations are also damped by this mechanism and planets tend to share a same orbital plane, which is consistent with {\it Kepler} data.
Such efficient collision damping is not found when we start calculations from densely packed orbits of the protoplanets.
If the protoplanets are initially in the mean-motion resonances, which corresponds to well separated orbits, the {\it in situ} accretion model well reproduces the features of eccentricities and inclinations of multiple super-Earths/Earth systems discovered by RV and {\it Kepler} surveys.

\end{abstract}



\section{INTRODUCTION}

%Observation eccentricities
About 60 close-in super-Earths ($\leq 30M_{\oplus}$, $M_{\oplus}$ is the mass of the Earth) have been discovered by the radial velocity method so far \footnote{http://exoplanets.org}.
Fabrycky et al. (2014) showed that the {\it Kepler} survey found more than 818 super-Earth-sized ($\leq 6R_{\oplus}$, $R_{\oplus}$ is the radius of the Earth) candidates in 333 multiple systems.
These planets can be formed either by 1) type I migration of the full-sized planets that have formed in outer regions (e.g., Cresswell \& Nelson 2006, Cossou et al., 2014),
2) {\it in situ} accretion of planetesimals that formed there or have migrated from outer regions due to aerodynamical gas drag (e.g., Raymond et al., 2007; Chiang \& Laughlin 2013),
3) {\it in situ} accretion of protoplanets that have migrated from outer regions due to type I migration (Terquem \& Papaloizou 2007; Ogihara \& Ida 2009; Ida \& Lin 2010).
Model 1 has a difficulty of why the full-sized planets were able to avoid runaway gas accretion.
In models 2 and 3, if the growth beyond a critical core mass occurs after disk gas depletion, the runaway gas accretion is avoided, while observed pile-ups near mean-motion resonances are not easy to be explained.
In model 2, accumulation of large amount of planetesimals may be a difficulty\footnote{
Note that Chiang \& Laughlin (2013) assumed that the feeding zone width is as large as orbital radius itself, while it is usually set to be several to ten Hill radii. 
So, much larger planetesimal surface density would actually be required than that estimated by Chiang \& Laughlin (2013).}.
In model 3, type I migration may be able to bring larger amount of solid materials to inner regions, as explained below, although the total mass of predicted super-Earths may still be smaller than the observed one (e.g., Ida \& Lin 2010, Ida et al., 2013). 
In model 3, giant impacts among protoplanets that have migrated from outer regions occur after disk depletion (giant impacts could also occur in model 2).
It is often considered that orbital eccentricities resulted in by giant impacts are larger than the observed values (see below), which is also a problem for model 3. 
In the present paper, however, we will show that eccentricities resulted in by giant impacts should be as small as the observed level.

The details of model 3 are as follows. 
Type I migration is halted at the disk inner edge and subsequently migrating protoplanets could be trapped by mean-motion resonances of preceding one stopped at the disk edge.
Terquem \& Papalouzou (2007) showed that type I migration is too fast for protoplanets to be captured by resonances at first encounters.
They are trapped in resonances after close scattering and coagulation near the disk edge, resulting in a few coagulated planets in resonant orbits.
Their orbits are stable even after disk depletion.
It is inconsistent with data of {\it Kepler} candidates that most of multiple systems are off-resonant, unless additional effects to deviate the planets from the resonant configurations are applied (e.g., Papaloizou 2011).
Model 1 also requires a similar process to realize off-resonant orbits (Goldreich \& Schlichting 2014).

Ogihara \& Ida (2009) found that if type I migration rate is considerably reduced from that derived by Tanaka et al. (2002), protoplanets are resonantly trapped at first encounters, resulting in resonant systems consisting of a large number of protoplanets.
In the presence of disk gas in which the eccentricities are strongly damped by planet-disk interactions (e.g., Tanaka \& Ward 2004), the protoplanets' orbits are stable.
However, if the number of resonant protoplanets exceeds a critical value, the systems can become unstable after disk gas depletion (Matsumoto et al. 2012).
The following orbit crossing and giant impacts result in off-resonant multiple super-Earth systems (Ogihara \& Ida 2009), which could be consistent with the off-resonant {\it Kepler} systems.

The observed orbital eccentricities and inclinations constrain the formation model as well as semimajor axis distributions.
Eccentricities are estimated by radial velocity measurements, while mutual inclinations are constrained in multiple systems by transit detection.
In model 1, the planetary orbits should be almost coplanar and circular. Higher eccentricities and inclinations are expected in models 2 and 3.
In this paper, we will discuss the latter model in comparison with the observation, because model 1 has a difficulty of runaway gas accretion and we will show that the eccentricities and inclinations in the latter models are not actually high.

While some gaseous giant planets have eccentricities as large as 0.9, super-Earths and Neptune-type planets generally have smaller eccentricities than gas giants (Mayor et al. 2011).
Scattering between similar-sized planets can excite their velocity dispersion up to their surface escape velocities ($v_{\rm esc}$) during their assemblage stages in a gas free conditions (e.g., Safronov 1969; Aarseth et al., 1993; Kokubo \& Ida 2002).
If the velocity dispersion exceeds $v_{\rm esc}$, the collision cross section becomes larger than the scattering cross section, so that the excitation of eccentricities is saturated at $v_{\rm esc}$.
The corresponding eccentricity is given by $e_{\rm esc} \sim v_{\rm esc}/ v_{\rm K} $ where $v_{\rm K}$ is the Kepler velocity, which we call ``escape eccentricity":
\begin{eqnarray}
 e_{\rm esc} &=& \sqrt{\frac{2(M_k+M_l)}{M_*} \frac{a}{R_k+R_l} } \nonumber\\
 &\simeq& 0.19 \left(\frac{M_k+M_l}{10M_{\oplus}} \right)^{1/3} \left( \frac{\rho}{3\ {\rm g cm^{-3}}} \right)^{1/6}\left( \frac{a}{0.1\ \rm AU} \right)^{1/2} \left( \frac{M_*}{M_{\odot} }\right)^{-1/2},
 \label{eq:e_esc}
\end{eqnarray}
where $M_*$ is the mass of the central star, $M_{\odot}$ is the solar mass, $M_k$ and $M_l$ are masses of protoplanets, $M_{\oplus}$ is the Earth mass, $R_k$ and $R_l$ are physical radii of protoplanets, $\rho$ is the material density of protoplanets, and $a$ is the semimajor axis of protoplanets (Kokubo \& Ida 2002).

\begin{figure}[htpb]
\begin{minipage}{0.8\hsize}
\includegraphics[angle=-90,width=.5\linewidth]{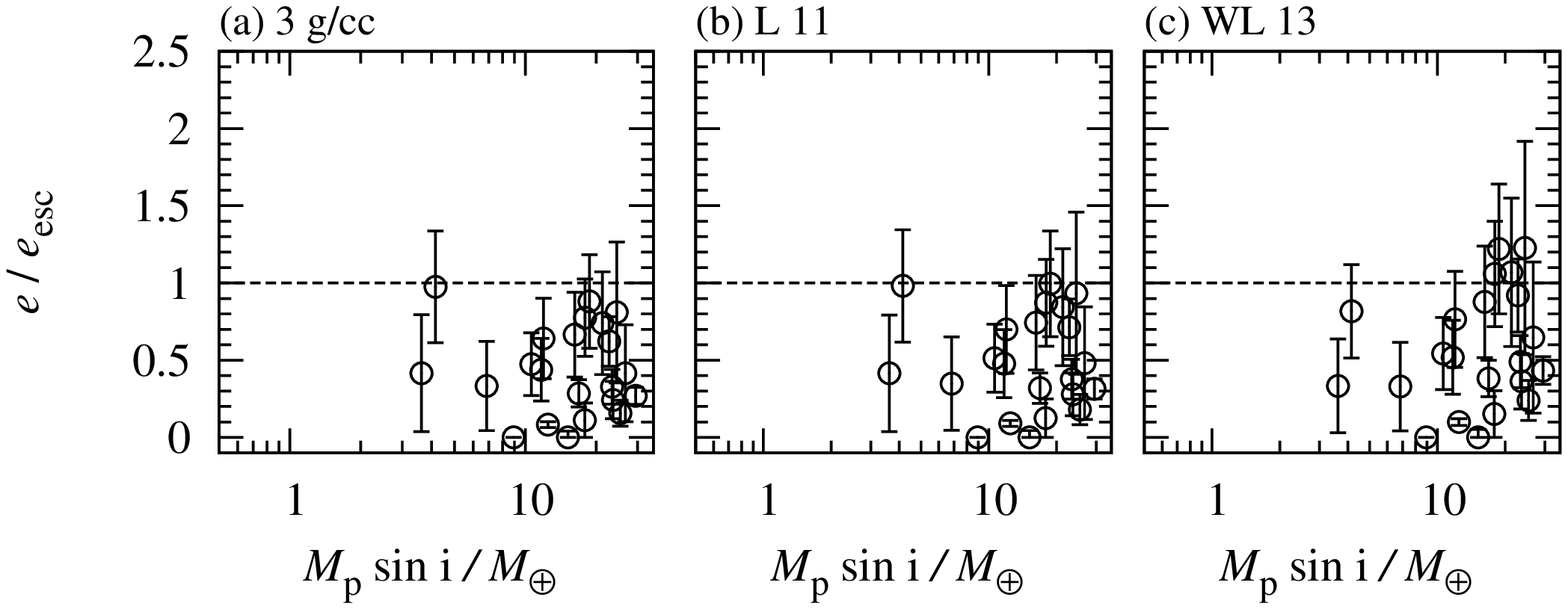}
\end{minipage}
\caption{\small{
 The exoplanets detected by RV method are plotted.
 Data was extracted from http://exoplanets.org.
 The horizontal axis is the planetary mass normalized by the Earth mass.
 The vertical axis is the eccentricities normalized by their escape eccentricities, defined by equation (\ref{eq:e_esc}). 
 We adopt different density models in each panel. 
 In panel (a), we adopt $\rho=3\ {\rm g/cm^3}$ for planets whose densities are not known. 
 In panel (b), planetary radii are given by $M_{\rm p} /M_{\oplus} = (R_{\rm p}/R_{\oplus})^{2.06}$, following Lissauer et al. (2011).
 In panel (c), planetary radii are given by $M_{\rm p} /M_{\oplus} = 3(R_{\rm p}/R_{\oplus})$, following Wu \& Lithwick (2013).
 The planets whose semimajor axes are greater than 0.1 AU are plotted.
 We remove planets whose eccentricities are unknown and assumed to be 0.
 The horizontal dashed line is $e/e_{\rm esc}=1$.
\newline}}
\label{fig:Me_e_esc_ob}
\end{figure}

However, most of observed super-Earths have smaller eccentricities than their escape eccentricities.
In Figure \ref{fig:Me_e_esc_ob}, we show observed eccentricities of planets observed by RV method scaled by their escape eccentricities, where we omit planets inside of 0.1 AU because their eccentricities can be damped by tidal interactions with their host stars (Goldreich \& Soter 1966).
Since planetary radius is needed to estimate $e_{\rm esc}$ (RV observations give only planetary minimum mass), we assume a mass-radius relationship for planets whose densities are not known. 
Figure \ref{fig:Me_e_esc_ob} shows $e/e_{\rm esc}$ of observed 22 planets as a function of planetary mass. 
Since the mass-radius relation has large uncertainty, we tested three models.
In panel (a), densities are given by 3 g/cm$^3$, independent on planetary mass. 
This is equivalent to the mass($M_{\rm p}$)-radius($R_{\rm p}$) relation, $M_{\rm p} /M_{\oplus}= 0.54 (R_{\rm p}/R_{\oplus})^3$, where $R_{\oplus}$ is the Earth radius.
This simple model is often used in N-body simulations for rocky planetesimals or protoplanets (e.g., Kokubo et al., 2006). 
This panel shows that super-Earths have smaller eccentricities than their escape eccentricities.
However, it is observationally suggested that larger super-Earths tend to have lower bulk densities.
Lissauer et al. (2011) derived an empirical mass-radius relation as $M_{\rm p}/M_{\oplus}=(R_{\rm p}/R_{\oplus})^{2.06}$ by fitting the planet in the solar system. 
With this mass-radius relation, the density of $30M_{\oplus}$ planet is 1.2 g/cm$^3$. 
In panel (b), this relation is used. 
Even with different mass-radius relations, the $e/e_{\rm esc}$ distribution in panel (b) is similar to that in panel (a).
Since the escape eccentricity is proportional to $\rho^{1/6}$, the different density does not significantly affect the result. 
From the lower density of observed larger exoplanets, Wu \& Lithwick (2013) and Weiss \& Marcy (2014) derived mass-radius relations with stronger density-dependence on mass. 
Wu \& Lithwick (2013) derived $M_{\rm p}/M_{\oplus}=3(R_{\rm p}/R_{\oplus})$ for $1.6\leq R_{\rm p}/R_{\oplus}<7$ planets (Weiss \& Marcy (2014) derived a similar relation).
This model is adopted in panel (c). 
Although 4 planets have $e>e_{\rm esc}$, they can be $e < e_{\rm esc}$ within the error bars. 
Furthermore, all of them have $M_{\rm p}\geq 18M_{\oplus}$ and their densities are estimated to be $\rho\lesssim0.4$ g/cm$^3$, which may be lower-estimated. 
Thereby, We conclude that the eccentricities of super-Earths are $e<e_{\rm esc}$ .
This means that super-Earths were formed through {\it in situ} coalescence followed by some eccentricity damping or not formed through the {\it in situ} coalescence.
Planetesimal accretion near 1 AU has been extensively studied.
Planetesimals grow through runaway growth in early stage (e.g., Wetherill \& Stewart 1989; Kokubo \& Ida 1996) and the oligarchic growth follows (e.g., Kokubo \& Ida 1998).
In these stages, protoplanets grow up, accreting surrounding planetesimals crossing their orbits from various directions.
Planetesimals prevent protoplanets from orbital crossings between protoplanets thanks to the dynamical friction.
This process is referred to as the orbital repulsion (Kokubo \& Ida 1995).
After most protoplanets accrete planetesimals in their feeding zones, the dynamical friction becomes no more available.
After disk gas is depleted, the planet-disk interaction is not available for the eccentricity damping as well.
Then, orbital instability occurs.
This stage is called as the giant impact stage.
The accretions of protoplanets in the giant impact stage are investigated in several papers (e.g., Chambers \& Wetherill 1998; Agnor et al., 1999; Raymond et al., 2004; Kokubo et al. 2006).
They are successful in reproducing terrestrial planets in the solar system in some aspects, such that the Earth and Venus mass planets are formed around 1 AU, if the range of initial semimajor axes of protoplanets is relatively restricted (Hansen 2009).

The eccentricities of the formed Earth-mass planets are usually $e\simeq 0.1$ (Kokubo et al., 2006).
While they are larger than the current free eccentricities of Venus and the Earth $\sim 0.02-0.03$, they are 3 times smaller than $e_{\rm esc}$.
Although some external damping mechanisms, such as dynamical friction from residual planetesimals (O'Brien et al. 2006) or planet-disk interaction (dynamical friction from disk gas; Kominami \& Ida 2004) need to be taken into account to reproduce the current eccentricities of Venus and the Earth, the low eccentricities of observed super-Earths can be potentially explained by the accretion through the giant impacts in gas-free environment.
We will make clear why eccentricities of formed planets become smaller than $e_{\rm esc}$.

For transit of multiple planets at $a < a_{\rm out}$ to be detected, mutual inclinations must be within $2.6 (R_*/R_{\odot}) (a/0.1\ {\rm AU})^{-1}$ degree.
Fabrycky et al. (2014) suggested that the typical mutual inclination of multiple super-Earths in {\it Kepler} candidates lies firmly in the range $i=1.0-2.2$ degree. 
On the other hand, numerical simulations show that close scattering between planetesimals realize $e\sim 2i$ (e.g., Ida \& Makino 1992; Shiidsuka \& Ida 1999).
For $e\sim e_{\rm esc}$, $i\sim i_{\rm esc} \sim e_{\rm esc}/2$, which is 5.4 degree for a $10M_{\oplus}$ super-Earth at 0.1AU.
This is larger than the mutual inclinations of {\it Kepler} systems.
However, N-body simulations show that Earth mass planets formed around 1 AU normally have $i\simeq 3$ degree (e.g., Kokubo et al. 2006), while $i_{\rm esc}\sim 8.6$ degree.
This result suggests that small inclinations of the {\it Kepler} systems can be realized by the {\it in situ} accretion in gas-free environment.

In this paper, we study how the final velocity dispersion becomes smaller than the escape velocity to account for the small eccentricities of observed super-Earths through $N$-body simulations of {\it in situ} accretion of planets in giant impact phase near the central star and analytical arguments.
In Section \ref{sect:model_II}, we outline the numerical methods and initial conditions of protoplanets.
Our results of $N$-body simulations are presented in Section \ref{sect:results_II}.
From 35 simulations in section \ref{sect:N_II}, we find the eccentricity and inclination of the largest planets are typically much smaller than the escape velocity. 
Because planetesimals and planet-disk interaction are not included in the simulation, the low eccentricities and inclinations are not due to dynamical friction.
In section \ref{sect:col_3}, we find collisions in the giant impact stage tend to occur when the angles between pericenters are around 180 degree and they are responsible for the low eccentricities and inclinations.
We summarize the results in Section \ref{sect:sum}.

\section{NUMERICAL MODEL}\label{sect:model_II}

We perform $N$-body simulations of the planet accretion starting from protoplanets without small planetesimals.
Orbital evolution of protoplanets is obtained by the numerical integration of
\begin{eqnarray}
 \frac{d^2 \textrm{\boldmath $r$}_i}{dt^2} = -{\rm G}M_{*}\frac{\textrm{\boldmath $r$}_i}{r_i^3} - \sum_{ j \neq i} {\rm G} M_{j} \frac{\textrm{\boldmath $r$}_{ ij} }{r_{ ij}^3} - \sum_j {\rm G} M_{j} \frac{\textrm{\boldmath $r$}_j}{r_j^3},
\end{eqnarray}
where ${\rm G}$ is the gravitational constant, $ \textrm{\boldmath $r$}_i$ are the coordinate of the $i$-th protoplanets numbered from the innermost, and $ \textrm{\boldmath $r$}_{ij}$ is the relative distance of the planets $i$ and $j$.
In our calculations, the central star has a solar mass.
The numerical scheme is the fourth-order Hermite scheme.
We assume perfect accretion, i.e., planets always accrete without bouncing nor fragmenting, and the total momentum is conserved at a collision.
Every time a collision is taken place, we record the orbital elements and masses of protoplanets before and after the collision.

We perform 2 sets of simulations: $5-16$ non-equal-mass protoplanets (which is hereafter referred to as ``$N$-body set") and three equal-mass protoplanets (``three-planet set").
In a standard case (case A; 20 runs) of the $N$-body set, we distribute protoplanets in a range from $0.05$ AU ($=a_1$) to 0.29 AU.
The number of the planets is $N=16$.
Their total mass is $M_{\rm tot}=17.3M_{\oplus}$.
Individual masses are given by
\begin{eqnarray}
 M \simeq 0.9 \left( \frac{\Sigma_{\rm 1}}{100\ {\rm g cm^{-2}}} \right)^{3/2} \left( \frac{a}{\rm 0.1\ AU} \right)^{3/4} M_{\oplus},
\end{eqnarray}
with $\Sigma_{\rm 1}=100 \ {\rm g cm^{-2}}$.
Although these protoplanets may have migrated from outer regions, we used a formula for isolation masses in {\it in situ} oligarchic growth (Kokubo \& Ida 2002).
Orbital separations ($b$) are $10r_{\rm H}$ where $r_{\rm H}$ is the Hill radius.
The initial individual masses do not affect the results as long as $b\sim 10r_{\rm H}$.
We also performed similar simulations with different $\Sigma_1$ and $N$ (accordingly, $M_{\rm tot}$ is also different) with the same $b$: case B, C, and D (5 runs for each set).
Planetary physical radii are calculated using a material density of $\rho=3\ {\rm g cm^{-3}}$.
The initial eccentricities and inclinations of protoplanets are given by the Rayleigh distribution.
The dispersions of eccentricity and inclination are $\langle e^2 \rangle^{1/2}=3\times10^{-2}(\Sigma_1/100\ {\rm gcm^{-2}})^{1/2}$ and $\langle i^2 \rangle^{1/2}=6\times10^{-4}(\Sigma_1/100\ {\rm gcm^{-2}})^{1/2}$ radian.
The initial conditions are summarized in Table \ref{table:cases_N}.
The simulations follow the evolution of protoplanet systems for $10^8$ Kepler time of the innermost planet. 
In some simulations, we calculate $3\times 10^8$ Kepler time and confirm that the resultant planet are stable.

In the next set of simulations, we perform many runs using equal-mass three-planets for statistical surveys.
These simple-settings enable us to control planetary masses and semimajor axes of colliding bodies.
In the three-planet cases, we give the semimajor axes of middle planets ($a_2$).
Inner planets and outer planets are set at $a_{1, 3} = a_2 \pm \tilde{b}r_{\rm H}$.
The Hill radius is given by $r_{\rm H}=(2M_{\rm p}/3M_*)^{1/3}a_2 \simeq 1.26\times10^{-3}(M_{\rm p}/M_{\oplus})^{1/3} (a_2/{\rm 0.1\ AU})$ AU.
The planetary radius ($R_{\rm p}$), planetary mass $(M_{\rm p})$, the semimajor axis of the middle planet ($a_2$), the orbital separations normalized by the Hill radius ($\tilde{b}$), and the initial eccentricities ($e_{\rm ini}$) are free parameters.
The orbits of planets are coplanar. 
We perform 11 cases in total, and we calculate 100 runs in each case changing initial orbital angles of the protoplanets randomly. %
The initial conditions are summarized in Table \ref{table:cases_3}.
We also calculate systems composed by non-zero inclination planets ($a_2i_{\rm ini}\leq2r_{\rm H}$), and confirm that collisions between inclined planets show the same tendency of collisions between planets in coplanar orbits, although the results in non-coplanar cases are not presented in this paper.

\section{RESULTS}\label{sect:results_II}

We first present the results of the $N$-body set ($N=5-16$).
We focus on the eccentricities and longitudes of pericenter before and after collisions to investigate eccentricity evolution through collisions.
Next, we show the results of the three-planet set, to study the dependences of eccentricity evolution on initial conditions.
Through these calculations, we explain intrinsic dynamics to cause the efficient collisional damping for the systems starting from moderately separated orbits.

\subsection{Results of $N$-body Set}\label{sect:N_II}

\begin{figure}[htpb]
\includegraphics[angle=-90,width=1.\linewidth]{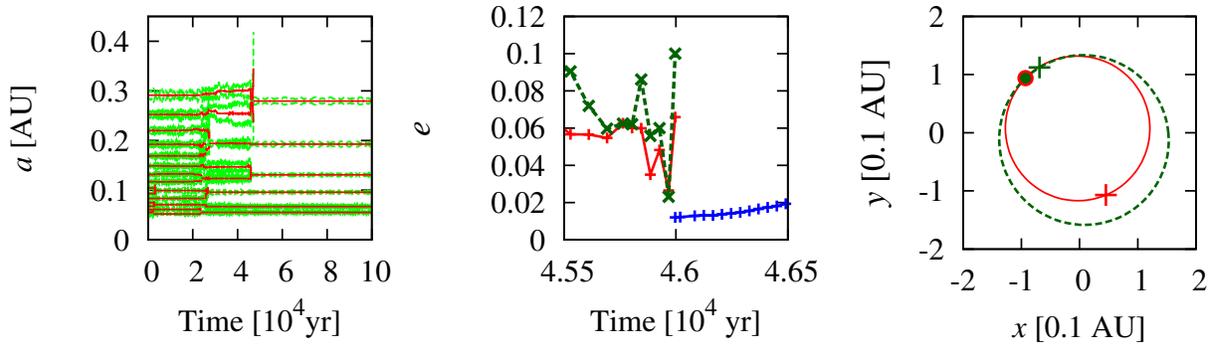}
\caption{\small{
The orbital evolution of a run of case A.
{\it Left}:
The time evolution of semimajor axes (solid lines), and pericenters and apocenters (dashed lines) of all planets.
{\it Middle }:
 Close-up of the left panel at a collision.
 The time evolution of the eccentricities of two colliding protoplanets is plotted.
 They collide at $t=4.60\times10^4$ yr.
 Red solid, green dashed, and blue solid lines are inner, outer, and merged protoplanets, respectively.
{\it Right}:
The face-on view of the orbits of colliding bodies is plotted.
The orbits of the inner and outer planet are shown in the solid curve and dashed curve, respectively.
The central star is located at the origin.
Circles indicate the positions of protoplanets.
Crosses are the locations of their pericenters.
The two pericenters are located in the opposite direction.
\newline}}
\label{fig:evolution_example}
\end{figure}

The typical orbital evolution is shown in the left penal of Figure \ref{fig:evolution_example}.
This figure shows time evolution of semimajor axes, pericenters, and apocenters of planets for $1.0\times 10^5$ yr.
In this calculation, six planets are finally formed.
They have final masses between $1.3 M_{\oplus}$ and $4.5 M_{\oplus}$.
Their escape eccentricities are $e_{\rm esc}= 0.079$ - 0.23 (equation (\ref{eq:e_esc})).
The final eccentricities of planets are between 0.015 and 0.047, which are much smaller than $e_{\rm esc}$.
The $\langle e \rangle / e_{\rm esc}$ of the largest, the second largest, and other planets in case A, B, C and D are summarized in Table \ref{table:results_N_1_2}.
In all cases, $\langle e \rangle / e_{\rm esc}$ is less than unity.
In particular, for the largest bodies, $\langle e \rangle / e_{\rm esc}$ is only 0.1-0.2 except case D in which $M_{\rm tot}$ is extremely small ($\sim0.15M_{\oplus}$).

%collision, e damping
These $e/e_{\rm esc}<1$ features are caused only by collisions.
The middle panel of Figure \ref{fig:evolution_example} is the closeup of the eccentricity evolution of protoplanets at a collision.
The fourth innermost planet with $2.1M_{\oplus}$ and the fifth one with $1.2M_{\oplus}$ collide at $t\simeq 4.6\times 10^4$ yr.
The eccentricities of the inner and outer planets are 0.066 and 0.10 just before the collision, which are comparable to $e_{\rm esc}\sim 0.1$.
However, the eccentricity of the merged body is 0.012 just after the collision, which is an order of magnitude smaller than $e_{\rm esc}$.

We show the orbits of two planets just before the collision in the right panel of Figure \ref{fig:evolution_example}.
The locations of the collision and their pericenters at the collision are shown by filled circles and crosses, respectively.
The azimuthal velocity at the apocenter of the inner planet is given by $\sqrt{GM_* (1-e_1)/a_1(1+e_1)} \simeq v_{\rm K} (1-e_1)$, where $v_{\rm K} $ is the Keplerian velocity at the collision location, while that at the pericenter of the outer planet is $v_{\rm K} (1+e_2)$.
Since they have similar masses and eccentricities, the velocity of the merged body should be $\sim v_{\rm K}$ due to conservation of total momentum, which means that the orbit of the merged body is nearly circular.
If the orbital separation is comparable to radial excursion due to the eccentricities, collisions occur only when the apocenter of the inner planet meets the pericenter of the outer planet (if their pericenters are aligned, their orbits never cross).
For such orbital separation, the collisional damping for eccentricity is always very efficient.
If the orbits of protoplanets are nearly circular and well separated, the eccentricities are excited only by secular perturbations.
The eccentricities are secularly increased until the apocenter distance of the inner planet approaches the pericenter distance of the outer planet and a collision between them occurs.

More detailed analysis can be done using Laplace-Runge-Lenz vector.
The mass-weighted total Lenz vector is conserved during scattering and even collisions under Hill's approximation (Nakazawa \& Ida 1988).
According to the conservation, when two planets $k$ and $l$ collide and are merged into a planet $kl$, the eccentricity of the merged body ($e_{kl}$) is written as
\begin{eqnarray}
 (M_k+M_l)^2 e_{kl}^2 = M_k^2 e_k^2 + M_l^2 e_l^2 + 2 M_k M_l e_k e_l \cos{(\varpi_k-\varpi_l)},
 \label{eq:LRL}
\end{eqnarray}
where $\varpi$ are longitudes of pericenters of the bodies.
When longitudes of pericenters of bodies are randomly distributed, the average of $e_{kl}^2$ becomes
\begin{eqnarray}
 (M_k+M_l)^2 e_{kl}^2 = M_k^2 e_k^2 + M_l^2 e_l^2.
 \label{eq:Ohtsuki1992}
\end{eqnarray}
This equation means that the eccentricity of the merged body is comparable with those of the colliding bodies;
in the case of $M_k=M_l$ and $e_k=e_l$, $e_{kl}=e_k/\sqrt{2}$.
The approximation of random $\varpi$ is valid if we consider a radially packed distribution of bodies.
The validity of equation (\ref{eq:Ohtsuki1992}) is confirmed by the $N$-body simulations for random velocity evolution of packed planetesimals (Ohtsuki 1992).

However, as already mentioned, in the case of giant impacts of protoplanets that have initially well separated orbits, $\Delta \varpi = \varpi_k-\varpi_l$ may be $\sim 180$ degree.
In this case, equation (\ref{eq:LRL}) implies
\begin{eqnarray}
 (M_k+M_l)^2 e_{kl}^2 \sim M_k^2 e_k^2 + M_l^2 e_l^2 - 2 M_k M_l e_k e_l.
 \label{eq:LRL_protoplanets}
\end{eqnarray}
When $M_k\sim M_l$, $e_{kl}^2\sim (e_k-e_l)^2/4$, which is much smaller than $e_{kl}^2\sim (e_k^2+e_l^2)/4$ given by equation (\ref{eq:Ohtsuki1992}).

For the collision in Figure \ref{fig:evolution_example}, the inner planet has $M_k=2.1M_{\oplus}$ and $e_k = 0.066$ and the outer planet has $M_l=1.2M_{\oplus}$ and $e_l = 0.10$.
We found $\Delta \varpi=171$ degree.
If we use equation (\ref{eq:Ohtsuki1992}), the estimated value of $e_{kl}$ is $\sim 0.056$.
However, it is $\sim 0.0080$ with equation (\ref{eq:LRL}), which is much more consistent with the orbital integration.

The collisions like the right panel of Figure \ref{fig:evolution_example} occur commonly between separated protoplanets.
In case A, there are 203 collisions in 20 runs.
The panel A of Figure \ref{fig:sort_varpi_diff_N} shows the distribution of $\Delta \varpi$ obtained in the 203 collisions.
It clearly shows that collisions tend to occur around $\Delta \varpi=180$ degree. 
Despite the difference in the number, masses, semimajor axes of protoplanets, similar peaks at $\Delta \varpi=180$ degree are found in the other cases.
The numbers of collisions are 13 in case B, 24 in case C, and 21 in case D.
The mean values and variances of $\Delta \varpi$ are $180\pm53$ degree in case A,
$177\pm30$ degree in case B,
$178\pm63$ degree in case C, and
$187\pm55$ degree in case D.

This means that the eccentricity of merged protoplanets tend to be much smaller than the estimation in equation (\ref{eq:Ohtsuki1992}).
This efficient eccentricity damping was mentioned in Raymond et al. (2006), although they did not analyzed the concentration of $\Delta \varpi$ around 180 degree.

If orbit crossing still continues, the damped eccentricity is excited again to $\sim e_{\rm esc}$.
However, because timescales ($\tau_{\rm cross}$) for orbital instability to start sensitively depend on the initial orbital separations (e.g., Chambers et al. 1996), $\tau_{\rm cross}$ of a system can jump up by several orders of magnitude at a collisional merging (see Figure 3 in Ida \& Lin 2010).
After that, the system becomes stable during main-sequence lifetime of the host stars and the damped eccentricities are remained.

We find inclinations are also significantly damped through collisions.
Figure \ref{fig:ik_ikl_ALL} shows the inclinations just before and after collisions.
The velocity component normal to the invariant plane depends on the ascending node. 
When a collision occurs at the ascending node and descending node of colliding bodies, the velocity component normal to the invariant plane of the merged body is much smaller than those of colliding bodies. 

\begin{figure}[htpb]
\includegraphics[angle=-90,width=.6\linewidth]{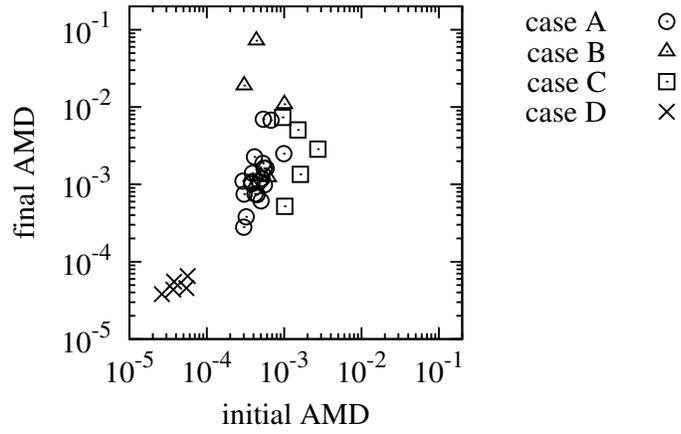}
\caption{\small 
The initial AMD and final AMD of calculations in $N$-body set are plotted. 
Circles are those in case A, triangles are in case B, squares are in case C, and crosses are in case D.
\newline}
\label{fig:amd_io_caseN}
\end{figure}

\begin{figure}[htpb]
\includegraphics[angle=-90,width=1.\linewidth]{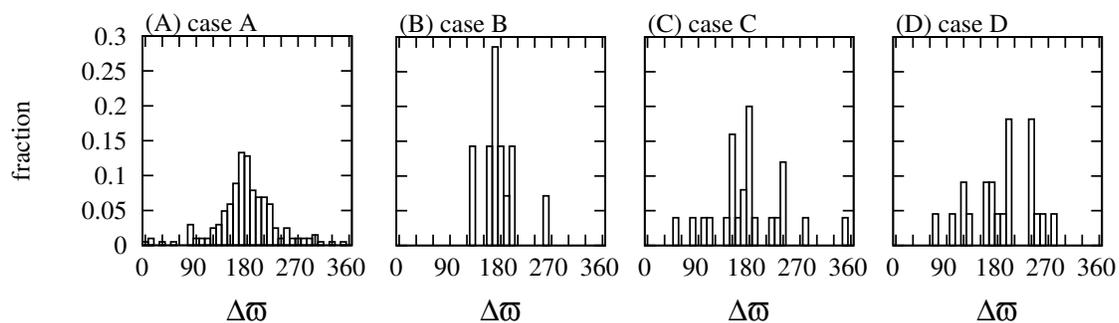}
\caption{\small{
The distributions of the difference between pericenters of colliding two bodies just before the collisions in $N$-body set are shown. 
Bin size is 10 degree.
In case A, There are 203 collisions in total.
Total numbers of collisions are 13-24, in the other cases.
The mean and variance of $\Delta \varpi$ are $180\pm 53$ degree in case A, 
$177\pm 30$ degree in case B, $178\pm 63$ degree in case C, and $187\pm 55$ degree in case D. 
\newline}}
\label{fig:sort_varpi_diff_N}
\end{figure}

\begin{figure}[htpb]
\includegraphics[angle=-90,width=.5\linewidth]{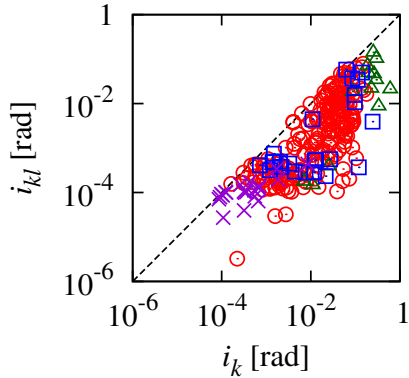}
\caption{\small{
Inclinations just before and after collisions in $N$-body cases are plotted in units of radian.
The larger inclination of the colliding two bodies is denoted as $i_k$, and $i_{kl}$ is the inclination of the merged body.
Circles are those in case A, triangles are in case B, squares are in case C, and crosses are in case D.
The dashed line is $i_{kl}=i_k$.
In every collision, $i_{kl}\leq i_k$ holds.
\newline}}
\label{fig:ik_ikl_ALL}
\end{figure}

The largest planets formed in case A have the averaged inclination of $\langle i\rangle=1.2\pm1.8$ degree.
Since the largest planets have a mean mass $\sim 4.4M_{\oplus}$ and semimajor axis $\sim0.2$ AU (Table \ref{table:results_N_1_2}), $i_{\rm esc}\sim e_{\rm esc}/2\simeq 6$ degree.
The inclinations of the largest planets are considerably smaller than $i_{\rm esc}$.
In other $N$-body set cases, formed planets also have smaller inclinations than $i_{\rm esc}$.
The inclinations of the largest bodies are $\langle i \rangle=4.7\pm1.9$ degree, $1.9\pm3.0$ degree, and $0.70\pm0.23$ degree in case B, case C, and case D, respectively.
Mutual inclinations in case A and C agree with those of {\it Kepler} planets, $i=1.0-2.2$ degree (Fabrycky et al. 2014).
Because of the inclination damping, the final planetary systems tend to be coplanar.
The means and variances of the typical mutual inclinations among all planets in a system are $\langle i_{\rm rel} \rangle=1.3\pm1.7$ degree, $4.4\pm7.0$ degree, $0.78\pm1.3$ degree, and $6.8\times10^{-3}\pm8.6\times10^{-3}$ degree in case A, case B, case C, and case D.
Because of larger semimajor axes of planets in case B, $i_{\rm esc}$ is larger.
Accordingly, $\langle i \rangle$ of the largest planets and $\langle i_{\rm rel} \rangle$ are larger in case B than those in the other cases, although $\langle i \rangle$ and $\langle i_{\rm rel} \rangle$ are still $< i_{\rm esc}$.
In general, angular momentum deficits (AMDs) are increased from initial values by scattering. 
However, the increase is not so significant except in case B (Figure \ref{fig:amd_io_caseN}).

In case A, C, and D, protoplanets tend to collide with the neighboring protoplanets rather than undergo global orbital instability.
In the proximity of their host stars, the ratio of Hill radii to physical radii is small, so that scatterings are less dominated over collisions than in outer regions.
In other words, $e_{\rm esc}$ and $i_{\rm esc}$ are smaller for smaller semimajor axis.
Furthermore, $e$ and $i$ are significantly smaller than $e_{\rm esc}$ and $i_{\rm esc}$.
Therefore, $e$ and $i$ can be very small through collisional damping in close-in regions.
Our results are not affected by the assumption of perfect accretion.
It was shown that the eccentricities in hybrid N-body and SPH simulations allowing collisional fragmentation by Kokubo \& Genda (2010) do not differ from those obtained in perfect accretion simulations.

\subsection{Results of Three-Planet Set}\label{sect:col_3}

The very effective eccentricity damping comes from the concentration of $\Delta \varpi$ on $180$ degree in collisions.
Let $\epsilon_{\varpi}=|\Delta \varpi-\pi|$. 
Assuming $\epsilon_{\varpi}\ll1$, equation (\ref{eq:LRL}) becomes
\begin{eqnarray}
 (M_k+M_l)^2 e_{kl}^2 = (M_k e_k - M_l e_l)^2 + M_k M_l e_k e_l \epsilon_{\varpi}^2,
 \label{eq:LRL_epsilon}
\end{eqnarray}
when $M_k=M_l$ and $e_k=e_l\sim e_{\rm esc}$,
\begin{eqnarray}
 \frac{e_{kl} }{e_{\rm esc} } \sim \frac{ \epsilon_{\varpi} }{2} \sim \frac{ \epsilon_{\varpi} }{115\ {\rm degree}},
 \label{eq:LRL_epsilon_2}
\end{eqnarray}
If $\epsilon_{\varpi} \lesssim$ 10 degree, the eccentricity of the merged body is an order of magnitude smaller than $e_{\rm esc}$.

In this section, we investigate how the concentration occurs through simpler three-planet simulations.
We change planetary radii, planetary masses, initial semimajor axes of the middle planets, initial orbital separations, and initial eccentricities.
The initial conditions are summarized in Table \ref{table:cases_3}.
The general features of results of three-planet calculations are basically similar to those of $N$-body set in the previous section.
The $\Delta \varpi$ distributions are peaked at $180$ degree, $e_{kl}$ is significantly smaller than $e_{\rm esc}$, and the estimated eccentricities in equation (\ref{eq:LRL}) agree with $e_{kl}$.

With initial spacing we use, the system in the three-planet set readily becomes unstable and orbit crossing starts.
In this case, the system enters stable state after a first collision and the eccentricity damped at the collision is not usually excited any more.
Thereby, $\Delta \varpi$, the degree of eccentricity damping and their dependences on initial conditions are better described than in $N$-body set.

\begin{figure}
\includegraphics[angle=-90,width=.5\linewidth]{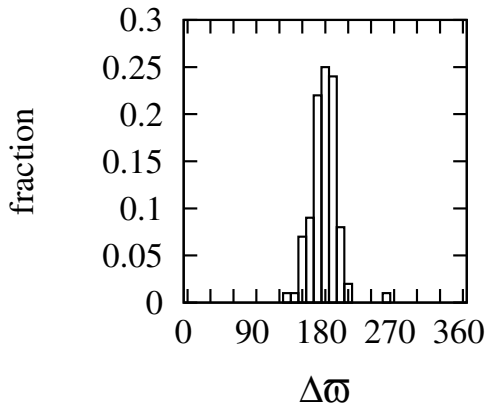}
\caption{
Same as Figure \ref{fig:sort_varpi_diff_N}, but for case 3A, where we calculate orbital evolution of three planets.
The mean and variance of $\Delta \varpi$ are $178\pm 17$ degree.
\newline}
\label{fig:3A}
\end{figure}

Case 3A is the standard case of three-planet calculations.
In this case, the second innermost planet is at 0.1 AU.
All planets have the same masses ($1M_{\oplus}$), radii ($1R_{\oplus}$), and initially circular orbits ($e=0$).
The initial orbital separations are given as $4r_{\rm H}$, which is equal to $5.04\times10^{-3}$ AU.
The distribution of differences between pericenters of colliding bodies ($\Delta \varpi$) in 100 simulations is shown in Figure \ref{fig:3A}.
The $\Delta \varpi$ distribution in case 3A is peaked at $180$ degree, in the same manner as case A (Figure \ref{fig:sort_varpi_diff_N}).
The variance of the $\Delta \varpi$ distribution ($\sigma_{\varpi}$) is 17 degree.

Typical evolution of eccentricities and arguments of pericenters in case 3A is as follows.
Eccentricities of planets increase by mutual scatterings, and collisions occur not long after their eccentricities exceed the eccentricities required for orbital crossing ($e_{\rm cross}= da/2a$).
When the orbits of two planets first become able to collide with each other, their orbits should have $e\sim e_{\rm cross}$.
However, the planets usually undergo close encounters and their $e$ are excited from $e_{\rm cross}$ before an actual collision.
In case 3A, the mean of the larger eccentricity between colliding planets before the collision is 0.094, which is larger than $e_{\rm cross}$($\simeq 0.028$), and comparable to $e_{\rm esc}$($\simeq 0.11$).
The detailed eccentricity evolution shows that colliding planets, which have typically $e=0.041\simeq 1.5e_{\rm cross}$ are pumped up to above eccentricity just before the collision.
The mean eccentricity after the collision is reduced to 0.015.
Substituting $\Delta \varpi=180\pm 17$ degree, $M_k=M_l$, and $e_k\sim e_l$ into equation (\ref{eq:LRL}), we get $e_{kl}/e_{k}=0.15$, which agrees well with the numerical value $0.015/0.094=0.16$.

In the following, we discuss $\Delta \varpi$ distribution using the pericenter dispersion $\sigma_{\varpi}$.
First, we analytically estimate $\sigma_{\varpi}$.
If eccentricities secularly increase from zero, the collision between two planets becomes possible when the apocenter of the inner planet ($Q_1$) contacts with a pericenter of the outer planet ($q_2$) with $\Delta \varpi=180$ degree.
In the following analysis, we neglect eccentricity excitation from $e_{\rm esc}$ for simplicity.
Although the assumption is not relevant enough, the analytical discussion neglecting the excitation well reproduces the numerical results.

\begin{figure}
\includegraphics[width=.5\linewidth]{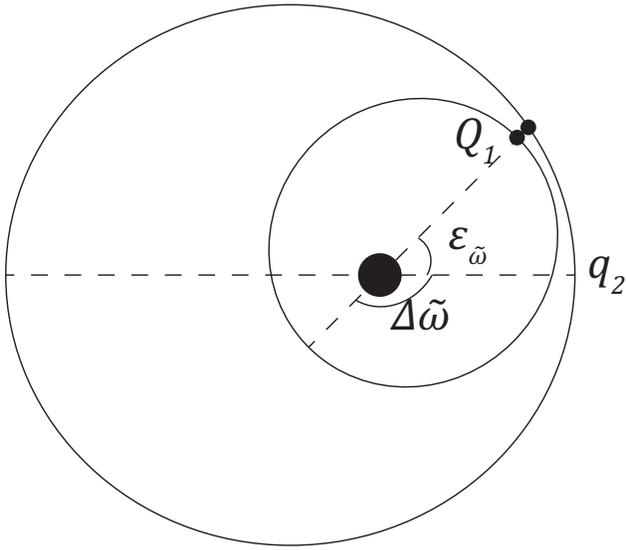}
\caption{
The orbital configuration of colliding planets is illustrated.
The large filled circle is the central star, and the small two filled circles are planets.
The two ellipses are the orbits of planets, and a dashed line is connecting the pericenter and apocenter of a planet.
The angle between the pericenters is $\Delta \varpi$.
Planets collide at the apocenter of the inner planet ($Q_1$), which is $\epsilon_{\varpi}$ rotated by the pericenter of the outer planets ($q_2$).
\newline}
\label{fig:collidable_angle}
\end{figure}

If we take into account physical radii of planets, the collisional point can rotate by an angle $\epsilon_{\varpi}$ from the pericenter of the outer planet (Figure \ref{fig:collidable_angle}).
In this case,
\begin{eqnarray}
\frac{a_2(1-e_2^2)}{1+e_2 \cos \epsilon_{\varpi}}-R_2 = a_1(1+e_1)+R_1,
\end{eqnarray}
where $R_1$ and $R_2$ are planetary physical radii of the inner and outer planets, respectively, and the true anomaly of the outer planet is equal to $\epsilon_{\varpi}$.
Under the assumption that planets collide at $Q_1$, the angle $\epsilon_{\varpi}+ \Delta \varpi=\pi$.
When we assume $Q_1 \simeq q_2 \gg R_{\rm tot}=R_1+R_2$,
\begin{eqnarray}
e_2 \cos \epsilon_{\varpi}= e_2-\frac{1+e_2}{1-e_2}\frac{R_{\rm tot}}{a_2}.
\label{eq:condition_col}
\end{eqnarray}
Assuming $e_2=e \ll 1$, $\epsilon_{\varpi} (\ll1$ radian) is
\begin{eqnarray}
\epsilon_{\varpi} \simeq \sqrt{\frac{2R_{\rm tot}}{ea_2}}.
\label{eq:eps_Rea}
\end{eqnarray}
With $e \simeq e_{\rm cross}= da/2a_2={\tilde b} r_{\rm H}/2a_2$, equation (\ref{eq:eps_Rea}) reads as
\begin{eqnarray}
\epsilon_{\varpi} &\sim& 2
\sqrt{\frac{2R_{\rm p}}{{\tilde b}r_{\rm H}}} \\
&=& 15 \left( \frac{R_{\rm p}}{R_{\oplus}}\right)^{1/2}
\left( \frac{M_{\rm p}}{M_{\oplus}}\right)^{-1/6}
\left( \frac{\tilde{b}}{4} \right)^{-1/2} \left( \frac{M_{*}}{M_{\odot}}\right)^{1/6}
\left( \frac{a}{0.1\ {\rm AU}} \right)^{-1/2} \ {\rm [degree]}.
\label{eq:epsomega}
\end{eqnarray}
The estimation of $\epsilon_{\varpi}$ agrees well with $\sigma_{\varpi}$ in case 3A ($\sigma_{\varpi}$=17 degree).
For collisions with $e>e_{\rm cross}$, simulated $\sigma_{\varpi}$ is larger than $\epsilon_{\varpi}$ given by equation (\ref{eq:epsomega}), resulting in inefficient eccentricity damping.
In the case of a dense orbital distribution of planetesimals, $\Delta \varpi$ is uniformly distributed without any concentration at $\pi$.
Then, the collisional damping is not effective and the eccentricities of merged bodies are similar to those during orbit crossing.

\begin{figure}
 \includegraphics[angle=-90,width=1.\linewidth]{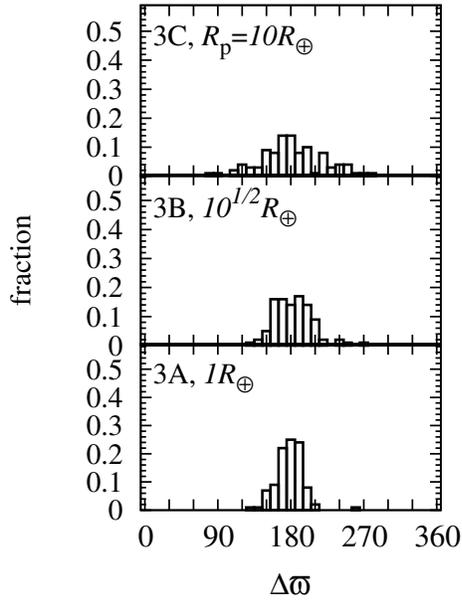}
\caption{\small{
Same as Figure \ref{fig:sort_varpi_diff_N}, but for case 3C, case 3B, and case 3A.
The panels are arranged in order of decreasing planetary radii from the top, $R_{\rm p}=10R_{\oplus}$ in case 3C, $10^{0.5}R_{\oplus}$ in case 3B, and $1R_{\oplus}$ in case 3A.
The variances of $\Delta \varpi$ are 39 degree in case 3C, 24 degree in case 3B, and 17 degree in case 3A.
Decreasing $R_{\rm p}$, $\Delta \varpi$ is more concentrated on 180 degree.
\newline}}
\label{fig:sort_varpi_diff_R}
\end{figure}

\begin{figure}
\includegraphics[angle=-90,width=1.\linewidth]{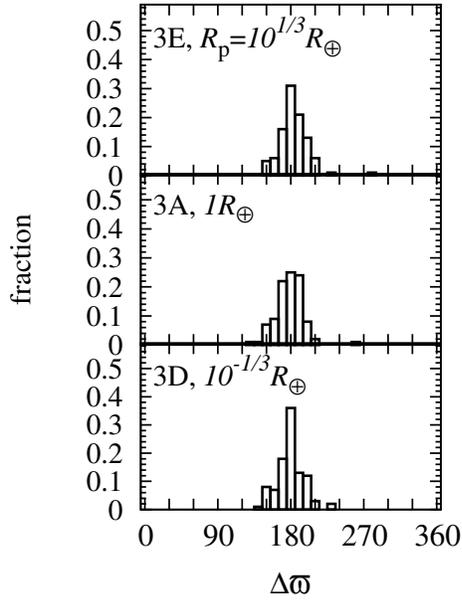}
\caption{\small{
Same as Figure \ref{fig:sort_varpi_diff_N}, but for case 3E, case 3A, and case 3D.
In these cases, we change planetary radii and masses keeping $\sqrt{R_{\rm p}/{\tilde b}r_{\rm H} }$ constant.
The panels are arranged in order of decreasing planetary radii from the top, $R_{\rm p}=10^{1/3}R_{\oplus}$ in case 3E, $1R_{\oplus}$ in case 3A, and $10^{-1/3}R_{\oplus}$ in case 3D.
The variances of $\Delta \varpi$ are 18 degree in case 3E, 17 degree in case 3A, and 16 degree in case 3D.
\newline}}
\label{fig:sort_MR}
\end{figure}

To check the validity of equation (\ref{eq:epsomega}), we perform additional runs.
Figure \ref{fig:sort_varpi_diff_R} shows the results of different $R_{\rm p}$.
We found $\sigma_{\varpi}=$17, 24, and 39 degree in case 3A ($R_{\rm p}=R_{\oplus}$), case 3B ($R_{\rm p}=10^{1/2}R_{\oplus}$), and case 3C ($R_{\rm p}=10R_{\oplus}$), respectively.
They agree with corresponding estimations, $\epsilon_{\varpi} \sim 15$, 27, and 47 degree.
We also perform calculations with changing planetary radii and masses in case 3D and case 3E, keeping $R_{\rm p}/da$ constant, which means that $\epsilon_{\varpi}$ is constant (Figure \ref{fig:sort_MR}).
The resultant $\sigma_{\varpi}$ is $16-18$ degree in all cases, while $R_{\rm p}$ changes a factor of 10.
These results also indicate that $M_{\rm p}$ affects $\sigma_{\varpi}$ through $da={\tilde b}r_{\rm H}$, and $\sigma_{\varpi}$ is proportional to $M_{\rm p}^{-1/6}$.

\begin{figure}
\includegraphics[angle=-90,width=1.\linewidth]{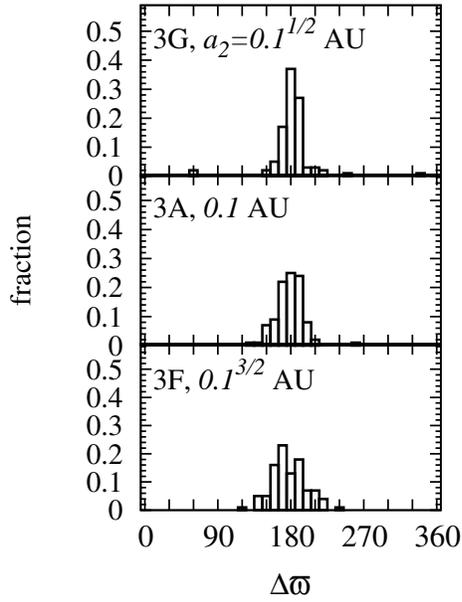}
\caption{\small{
Same as Figure \ref{fig:sort_varpi_diff_N}, but for case 3G, case 3A, and case 3F.
The panels are arranged in order of decreasing semimajor axis from the top, $0.1^{1/2}$ AU in case 3G, 0.1 AU in case 3A, and $a_2=0.1^{3/2}$ AU in case 3F.
The variances of $\Delta \varpi$ are 18 degree in case 3G, 17 degree in case 3A, and 22 degree in case 3F.
\newline}}
\label{fig:sort_a}
\end{figure}

\begin{figure}
\includegraphics[angle=-90,width=1.\linewidth]{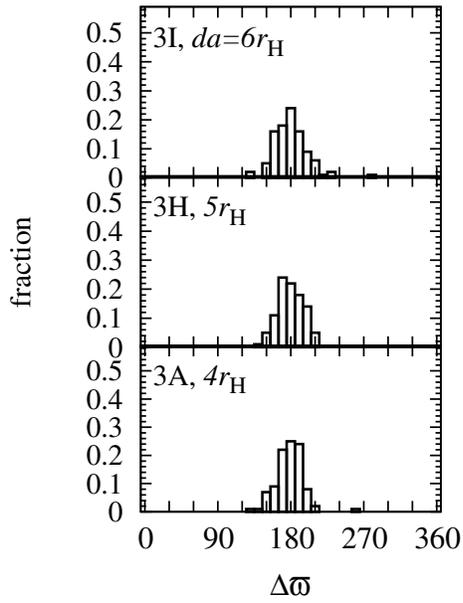}
\caption{\small{
Same as Figure \ref{fig:sort_varpi_diff_N}, but for case 3I, case 3H, and case 3A.
The panels are arranged in order of decreasing $da$ from the top, $da=6r_{\rm H}$ in case 3I, $5r_{\rm H}$ in case 3H, and $4r_{\rm H}$ in case 3A.
The variances of $\Delta \varpi$ are 21 degree in case 3I, 16 degree in case 3H, and 17 degree in case 3A.
\newline}}
\label{fig:sort_b}
\end{figure}

Next, we change $da$ by changing $a_2$ and $b$ with fixed $M_{\rm p}$.
The dependence on $a_2$ is shown in Figure \ref{fig:sort_a}.
For $a_2=0.1^{3/2}$ AU (case 3F), 1 AU (case 3A) and $0.1^{1/2}$ AU (case 3G), $\sigma_{\varpi}$ obtained by simulations are 22, 18 and 17 degree.
The estimated $\epsilon_{\varpi}$ by equation (\ref{eq:epsomega}) is not relevant enough in case 3G, because eccentricities are more highly pumped up before collisions.
Figure \ref{fig:sort_b} shows the results with ${\tilde b}=4$ (case 3A), 5 (case 3H) and 6 (case 3I).
Although the timescale for orbital instability to start is very different (e.g., Chambers et al. 1996), $\sigma_{\varpi}$ are similar: 17 degree in case 3A (${\tilde b}=4$), 16 degree in case 3H (${\tilde b}=5$), 21 degree in case 3I (${\tilde b}=6$), which are consistent with $\epsilon_{\varpi}\sim 15({\tilde b}/4)^{-1/2}$ degree (equation (\ref{eq:epsomega})).
These results show that concentrations of $\Delta \varpi$ at $\pi$ with $\sigma_{\varpi}\simeq 20$ degree is quite common as long as close-in regions ($\lesssim0.3$ AU) are considered.

\begin{figure}
\includegraphics[angle=-90,width=1.\linewidth]{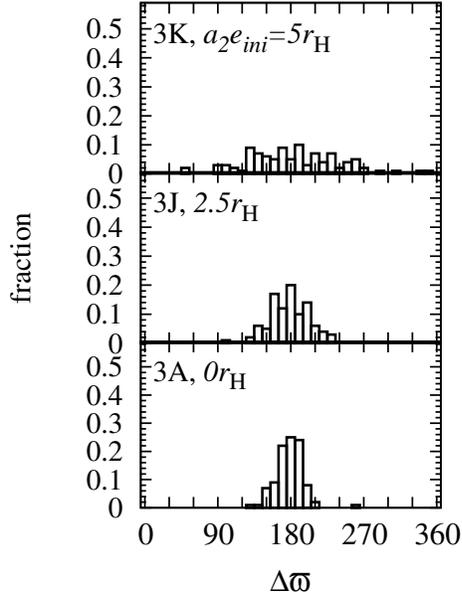}
\caption{\small{
Same as Figure \ref{fig:sort_varpi_diff_N}, but for case 3K, case 3J, and case 3A.
The panels are arranged in order of decreasing $a_2 e_{\rm ini}$ from the top, $a_2 e_{\rm ini}=5r_{\rm H}$ in case 3K, $2.5r_{\rm H}$ in case 3J, and $0r_{\rm H}$ in case 3A.
The variances of $\Delta \varpi$ are 56 degree in case 3K, 25 degree in case 3J, and 17 degree in case 3A.
\newline}}
\label{fig:sort_varpi_diff_e}
\end{figure}

The above results are applied to the systems in which mean orbital separations are larger than $2\sqrt{3}r_{\rm H}$ and orbital crossing doe not occur until eccentricities are gradually increased by distant perturbation.
In a system of equal-mass bodies with surface density $\Sigma$ at $a$, the mean orbital separation is
\begin{eqnarray}
 \tilde{b} \simeq 14\left(\frac{M}{M_{\oplus}}\right)^{2/3} \left(\frac{\Sigma}{3\times10^3\ {\rm gcm^{-2}} }\right)^{-1}
 \left(\frac{a}{\rm 0.1\ AU}\right)^{-2}.
\end{eqnarray}
If we consider early stages in which the systems consist of plenty of small planetesimals, ${\tilde b}$ is far smaller than $2\sqrt{3}$.
Then, the concentration of $\Delta \varpi$ does not occur and the collisional damping should be weak.
If $e$ is set such that the radial excursions are larger than orbital separations ($ea_2>{\tilde b}r_{\rm H}$), a situation is similar.
We set $ea_2>{\tilde b}r_{\rm H}$ in case 3K.
As expected, we obtain a relatively large value of $\sigma_{\varpi}$ (=56 degree) in this case, because collisions occur regardless of the directions of pericenters and $\Delta \varpi$ is no longer concentrated (Figure \ref{fig:sort_varpi_diff_e}).

\section{SUMMARY AND DISCUSSION}\label{sect:sum}
We have investigated the eccentricity damping through the giant impacts of the protoplanets in the proximity of the host stars.
First, we performed 20 runs of $N$-body simulations of protoplanets starting from 16 bodies of about Earth-mass at $0.05-0.29$ AU with orbital separation of 10 Hill radii (``$N$-body set").
We also performed simulations of runs with more and less massive bodies and runs at larger semimajor axes.
We have confirmed that eccentricities of formed planets are significantly lower than the eccentricity corresponding to velocity dispersion of their surface escape velocity (``escape" eccentricities $e_{\rm esc}$; see Section \ref{sect:N_II}).
For an Earth-mass body at 0.1AU, $e_{\rm esc}\sim0.1$.
During orbital crossing, eccentricities increase due to the mutual scatterings among protoplanets and reaches $\sim e_{\rm esc}$.
However, the eccentricities are damped by an order of magnitude at a collision.
When the orbits of the protoplanets are relatively separated and their eccentricities are secularly increased, the differences between the pericenters of colliding planets tend to be $\Delta \varpi\sim 180$ degree, i.e., the collisions occur at the apocenter of the inner body and the pericenter of the outer body.
Since the azimuthal velocity of the inner body is slower than the local Keplerian velocity and that of the outer body is faster and the two bodies have similar masses, the velocity of the merged body should be close to the local Keplerian velocity, which means that the orbit of the merged body is nearly circular.
We also described more detailed discussion on why the collision damping is so efficient, using conservation of Lenz vector in Hill's approximation.

The damped eccentricities are again excited up to $\sim e_{\rm esc}$ if orbital crossing continues.
However, after some merging, the planets become isolated from one another and the planets are remained in stable orbits with $e\ll e_{\rm esc}$.
We found that the inclinations of protoplanets are also damped through collisions.
The mutual inclinations among formed planets in the massive systems in close-in regions (case A and C) are $\langle i \rangle=1.3\pm1.7$ degree in case A and $\langle i \rangle=0.78\pm1.3$ degree in case C, respectively, which agree with those of observed super-Earths, $i=1.0-2.2$ degree (Fabrycky et al. 2014).

Next, we performed three-planet calculations (``Three-planet set").
With these simple systems, more detailed analysis on the collision damping can be done.
The eccentricity of a merged body is given approximately by $e\sim (\epsilon_{\varpi}/115\ {\rm degree})e_{\rm esc}$ where $\epsilon_{\varpi}$ is the width of the concentration of $\Delta \varpi$ around 180 degree (equation (\ref{eq:LRL_epsilon_2})).
If the concentration is high ($\epsilon_{\varpi}$ is small), $e$ can be much smaller than $e_{\rm esc}$.
Through the analytical argument and the results of the three-planet calculations, we found that $\epsilon_{\varpi}$ is approximated as $\epsilon_{\varpi}\sim 2(2R_{\rm p}/da)^{1/2}$ where $R_{\rm p}$ is planetary physical radius and $da$ is initial orbital separation.
That is, $e\sim 0.13(R_{\rm p}/R_{\oplus})^{1/2}(da/0.05 \ {\rm AU})^{-1/2} e_{\rm esc}$.
Note that in runaway and oligarchic stage, the collision damping is weak and $e\sim e_{\rm esc}$, because $da$ is very small and accordingly $\Delta \varpi$ is uniformly distributed.

Volk \& Gladman (2015) studied orbital stability of multiple super-Earth systems analogous to the systems discovered by {\it Kepler} and found that in some systems, global instability occurred and the collision velocities are larger than $2 v_{\rm esc}$ (the collisions are disruptive), while in our simulations, no global instability occurred and collision velocities are smaller than $1.5 v_{\rm esc}$. 
Although Volk \& Gladman (2015) did not explain the conditions for global instability, initial planetary mass distribution may have caused the difference. 
In their calculations, the maximum mass ratios among initial planets are generally much larger than ours. 
If a small-mass body collides with another small-mass one after scattering by a massive planet, the collision velocity can be larger than $v_{\rm esc}$ of the small bodies. 
While the results of Volk \& Gladman (2015) can be applied to some of Kepler systems, our results can be applied to other systems. 
Detailed investigation of scattering in systems with initially large mass ratios is left for future work.

\begin{figure}[htpb]
\includegraphics[angle=-90,width=.5\linewidth]{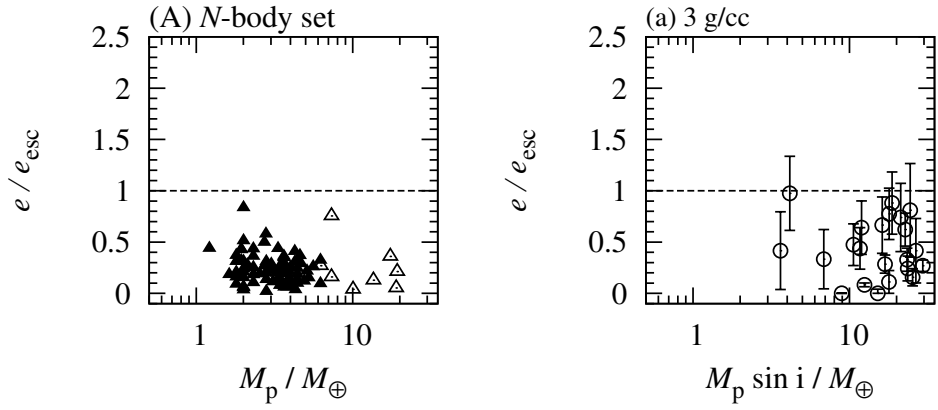}
\caption{\small{
(A):
Eccentricities of planets formed in our $N$-body calculations.
The formed planets that have $a>0.1$ AU in the results of case A and case C are plotted to compare to Figure \ref{fig:Me_e_esc_ob}.
The filled triangles are the planets in case A, and the open triangles are those in case C.
(a): Eccentricities normalized by their escape eccentricities with $\rho=3\ {\rm g/cm^3}$ of observed planets are plotted (Same as Figure \ref{fig:Me_e_esc_ob} a).
\newline}}
\label{fig:Me_e_esc_calc}
\end{figure}

Figure \ref{fig:Me_e_esc_calc}A shows the eccentricities $e$ and masses $M_{\rm p}$ of the formed planets in $N$-body set (cases A and C).
Observed data of super-Earths in Figure \ref{fig:Me_e_esc_ob} (a) are also plotted for comparison (Figure \ref{fig:Me_e_esc_calc}a).
We only plot planets at $a > 0.1$ AU, because eccentricities of planets at $a < 0.1$ AU may be damped by tidal dissipation.
The eccentricities of the formed super-Earths in our calculations are $e<0.5e_{\rm esc}$ except a few planets.
In the observed data, when we adopt $\rho=3$ g/cm$^3$, all planets have $e< e_{\rm esc}$.
The data are consistent with our result, although our results show slightly smaller $e$. 
Even if we use a more realistic mass-radius relation based on the Solar system planets derived by Lissauer et al. (2011), the result is hardly changed. 
With the mass-radius relation of Wu \& Lithwick (2013) that produces lower bulk density, $e > e_{\rm esc}$ for planets with $\gtrsim 20 M_{\oplus}$. 
However, $\rho \sim 0.4$ g/cm$^3$ for these planets may be too small. 
Furthermore, the best-fit eccentricities of planets detected with the low signal-to-noise ratio and the small number of observations by RV surveys tend to be larger than the true values (Shen \& Turner 2008). 
Therefore, we conclude that the eccentricities of observed close-in super-Earths are not inconsistent with our results.
Note that our results with $\rho=3$ g/cm$^3$ are equivalent to the results with other density and semimajor axis according to a scaling. 
The dynamical process is scaled by the ratio of the geometrical cross section to the scattering cross section with radius $r_{\rm H}$, which is proportional to $\rho^{-2/3}a^{-2}$. 
So, our results at $a=0.5$ AU correspond to the case of $\rho=1$ g/cm$^3$ and $a=0.035$ AU. 
For example, GJ 667C c has $e=0.97e_{\rm esc}$ and $4.2M_{\oplus}$. 
GJ 667C is a member of a triple stellar system.
The eccentricity of GJ 677C c could be affected by GJ 667A and GJ 667B.
Therefore, the eccentricities of super-Earths in observed data may be consistent with those obtained in our N-body simulations.

As we described in section 1, three models have been proposed for formation of close-in super-Earths: 1) type I migration of the full-sized planets that have formed in outer regions, 2) {\it in situ} accretion of planetesimals that formed there or have migrated from outer regions due to aerodynamical gas drag, and 3) {\it in situ} accretion of protoplanets that have migrated from outer regions due to type I migration.
As explained in Introduction, none of these can completely explain the presently known super-Earth systems.
What we argued in the present paper is that relatively low eccentricity found in {\it Kepler} systems is not a negative factor for a formation model of close-in super-Earth systems via giant impacts of protoplanets. 
With this result, model 3 might look promising. 
However, more detailed discussions on different aspects are needed to clarify the origin of super-Earth systems discovered by {\it Kepler}.

%
%
%Acknowledgement
\section*{}
We thank David Minton for comments that helped us improve the manuscript.
This research was supported by a grant for the Global COE Program, ¡ÉFrom the Earth to ¡ÉEarths¡É¡É, MEXT, Japan and a grant for JSPS (23103005) Grant-in-aid for Scientific Research on Innovative Areas.
Numerical computations were in part carried out on PC cluster at Center for Computational Astrophysics, National Astronomical Observatory of Japan.

\clearpage

\clearpage
\begin{deluxetable}{cccccccc}
\tabletypesize{\small}
\tablecaption{Initial conditions of $N$-body simulations\label{table:cases_N}}
\tablenum{1}
\renewcommand{\arraystretch}{0.8}
\tablehead{ \colhead{} & \colhead{$N$} & \colhead{$\Sigma_1$ } & \colhead{$a_{\rm 1}$} & \colhead{$M_{\rm tot}/M_{\oplus}$} & \colhead{$\langle e^2 \rangle^{1/2}$} &\colhead{$R_1/R_{\oplus}$} &\colhead{ \# of run}\\
\colhead{} & \colhead{} & \colhead{${\rm [gcm^{-2}]}$} & \colhead{[AU]} & \colhead{} & \colhead{} &\colhead{} &\colhead{} }
\startdata
 case A & 16 & 100 & 0.05& 17.3 & $3.16\times10^{-2}$ & 0.99 & 20\\%run A
 case B & 5 & 100 & 0.68& 25.3 & $3.16\times10^{-2}$ & 1.90 & 5\\%run C
 case C & 8 & 300 & 0.05& 37.2 & $5.48\times10^{-2}$ &1.72 &5 \\%run E
 case D & 8 & 10 & 0.05& 0.148 & $1.00\times10^{-2}$ & 0.31 &5 %run F
\enddata
\tablecomments{\small{
We give the number of protoplanets ($N$), the surface density of protoplanets at 1 AU ($\Sigma_1$), and the semimajor axis of the innermost protoplanet ($a_1$).
The total mass of a system ($M_{\rm tot}$), the dispersion of the eccentricity distribution ($\langle e^2 \rangle^{1/2}$), and the physical radius of the innermost planet ($R_1/R_{\oplus}$) are given from above parameters.
The last line of this table is the number of calculations in each $N$-body case.
\newline}}
\end{deluxetable}

\begin{deluxetable}{ccccccc}
\tabletypesize{\small}
\tablecaption{Initial conditions of three-planet simulations\label{table:cases_3}}
\tablenum{2}
\renewcommand{\arraystretch}{0.8}
\tablehead{ \colhead{}& \colhead{$R_{\rm p}/R_{\oplus}$} & \colhead{$M_{\rm p}/M_{\oplus}$} & \colhead{$a_2$} & \colhead{${\tilde b}$} & \colhead{$a_2 e_{\rm ini}/r_{\rm H}$} & \colhead{$da$} \\
\colhead{}& \colhead{} & \colhead{} & \colhead{[AU]} & \colhead{} & \colhead{} & \colhead{[AU]} }
\startdata
 case 3A & 1 & 1 & 0.1 & 4 & 0 & $5.04\times10^{-3}$ \\
 case 3B & $10^{1/2}$ & 1 & 0.1 & 4 & 0 & $5.04\times10^{-3}$ \\
 case 3C & 10 & 1 & 0.1 & 4 & 0 & $5.04\times10^{-3}$ \\
 case 3D & $10^{-1/3}$ & 0.1 & 0.1 & 4 & 0 & $2.34\times10^{-3}$ \\
 case 3E & $10^{1/3}$ & 10 & 0.1 & 4 & 0 & $1.09\times10^{-2}$ \\
 case 3F & 1 & 1 & $0.1^{3/2}$ & 4 & 0 & $5.04\times10^{-4}$ \\
 case 3G & 1 & 1 & $0.1^{1/2}$ & 4 & 0 & $1.59\times10^{-3}$ \\
 case 3H & 1 & 1 & 0.1 & 5 & 0 & $6.30\times10^{-3}$ \\
 case 3I & 1 & 1 & 0.1 & 6 & 0 & $7.56\times10^{-3}$ \\
 case 3J & 1 & 1 & 0.1 & 4 & 2.5 & $5.04\times10^{-3}$ \\
 case 3K & 1 & 1 & 0.1 & 4 & 5 & $5.04\times10^{-3}$
\enddata
 \tablecomments{\small{
 The radii of planets ($R_{\rm p}/R_{\oplus}$), the mass of planets ($M_{\rm p}/M_{\oplus}$), the semimajor axis of the middle planet ($a_2$), the orbital separations normalized by the Hill radius (${\tilde b}$), the initial eccentricities ($a_2 e_{\rm ini}/r_{\rm H}$) are given as initial conditions.
 The orbital separations between planets ($da$) is derived from $da={\tilde b}r_{\rm H}={\tilde b}(2M_{\rm p}/3M_*)^{1/3}a_2$.
 \newline}}
\end{deluxetable}

\begin{deluxetable}{@{\hspace{-1cm}}cccccccccccc}
\tabletypesize{\scriptsize}
\tablecaption{Results of $N$-body Cases\label{table:results_N_1_2}}
\tablenum{3}
\renewcommand{\arraystretch}{0.8}
\tablehead{\colhead{} & \colhead{$\langle n\rangle$} & \colhead{$\langle M_{l1}\rangle/M_{\oplus}$} & \colhead{ $\langle a_{l1}\rangle$} & \colhead{$\langle e_{\rm l1}\rangle/e_{\rm esc}$} & \colhead{$\langle M_{l2}\rangle/M_{\oplus}$} & \colhead{$\langle a_{l2}\rangle$} & \colhead{$\langle e_{\rm l2}\rangle/e_{\rm esc}$} & \colhead{$\langle e_{\rm rem}\rangle/e_{\rm esc}$}\\
\colhead{} & \colhead{} & \colhead{} & \colhead{[AU]} & \colhead{} & \colhead{} & \colhead{[AU]} & \colhead{} & \colhead{} }
\startdata
 case A & $5.85\pm1.01$ & $4.41\pm0.85$ & $0.20\pm0.05$ &$0.18\pm0.08$ & $3.72\pm0.66$ & $0.17\pm0.06$ & $0.24\pm0.26$ &$0.40\pm0.32$ \\%run A
 case B &$2.20\pm0.40$& $16.7\pm2.8$ & $1.05\pm0.23$ & $0.16\pm0.14$ & $7.10\pm2.4$ & $1.04\pm0.56$ & $0.41\pm0.35$ & $0.33\pm0.0060$\\%run C
 case C & $3.20\pm0.98$& $17.2\pm3.6$ & $0.11\pm0.03$ & $0.16\pm0.32$ & $11.7\pm3.4$ & $0.11\pm0.05$ & $0.32\pm0.20$ & $0.62\pm 0.53$\\%run E
 case D & $3.80\pm0.75$& $0.0504\pm0.0092$ & $0.060\pm0.002$ & $0.47\pm0.37$ & $0.0441\pm0.0076$ & $0.057\pm0.003$ &$0.36\pm0.10$ & $0.44\pm 0.23$%run F
 \enddata
 \tablecomments{\small{
 Average values of the numbers of formed planets ($\langle n\rangle$), the masses, semimajor axes, eccentricities normalized by their escape eccentricities of the largest planets ($\langle M_{l1}\rangle$, $\langle a_{l1}\rangle$, $\langle e_{\rm l1}\rangle/e_{\rm esc}$), and thoses of the second largest planets ($\langle M_{l2}\rangle$, $\langle a_{l2}\rangle$, $\langle e_{\rm l2}\rangle/e_{\rm esc}$) and the eccentricities normalized by their escape eccentricities of the other planets ($\langle e_{\rm rem}\rangle/e_{\rm esc}$) are denoted.
\newline
 }}
\end{deluxetable}

\end{document}